\documentclass[epsf]{aa}
\usepackage{graphicx}
\begin{document}
\def\cd{cd$^{-1}$}
\def\cds{cd$^{-1}$\,}
\def\pdu{$\phi_{21}$}
\def\ptu{$\phi_{31}$}
\def\pqu{$\phi_{41}$}
\def\Rtu{$R_{31}$}
\def\Rdu{$R_{21}$}
\def\fu{$f_1$}
\def\fd{$f_2$}
\def\ft{$f_3$}
\def\kms{km~s$^{-1}$}
\def\kmss{km~s$^{-1}$\,}
\title{Fourier decomposition and frequency analysis of the pulsating stars with
$P<$ 1 d in the OGLE database.}
\subtitle{I. Monoperiodic Delta Scuti, RRc and RRab variables. Separation
criteria and particularities\thanks{Tables 3, 4 and 5 are only available
in electronic form at the CDS via anonymous ftp to {\tt
cdsarc.u-strasbg.fr (130.79.128.5)} or via 
{\tt http://cdsweb.u-strasbg.fri/cgi-bin/qcat?J/A+A/.../...}}}
\author{E. Poretti}
\offprints{E. Poretti}
\institute{Osservatorio Astronomico di Brera, Via E.~Bianchi 46,
       I--23807 Merate, Italy \\
\email{poretti@merate.mi.astro.it} }
\date{Received 26 January 2001/~Accepted 14 March 2001}

\abstract{
The OGLE database is revisited to investigate in more detail 
the properties of the Fourier parameters. Methodological improvements
led us to identify a clear separation among High--Amplitude $\delta$ Scuti 
(HADS), RRc and RRab stars.  The bimodal
distribution of the \Rdu parameter in HADS stars is explained as a contamination
effect from RRc stars: there is evidence that all stars with
0.20$<P<$0.25 d are RRc variables. The previously claimed
existence of a subclass of unusual HADS is demonstrated to be a spurious
result. Candidate overtone pulsators are found among HADS and RRc variables. 
The properties of the Fourier parameters are discussed as a function of
the physical conditions in the stars involved. Among the field RRab stars we detected 
different light--curve groups producing distinct
``tails" in the Fourier plots for $P>$0.55~d;  evolutionary
phases or the combination of different physical conditions (not only metallicity)
are suggested to explain this  separation, observed also in the cluster
RRab stars. The stellar parameters of RRc stars in a given globular
cluster show different tendencies than those of RRc stars from different
clusters. 
\keywords
{Methods: data analysis -- Techniques: photometric --  
Astronomical data bases: miscellaneous -- Stars: oscillations --
Stars: variables: $\delta$ Sct -- Stars: variables: RR Lyr }
}
\authorrunning{E. Poretti}
\titlerunning{Monoperiodic stars in the OGLE database}
\maketitle

\section{Introduction}

The {\it Optical Gravitational Lensing Experiment} (OGLE; Udalski et al.
1993) is a project whose main goal is a systematic search for microlensing 
events. As a valuable by--product, a large amount of photometric data was
acquired on a variety of variable stars.
As demonstrated by several studies, the light curves of pulsating stars
depend on their physical parameters. Moreover, the observed light
curves can be compared with those derived from hydrodynamic
pulsational models. Therefore, once an observable quantity 
has been recognized as depending on stellar parameters, it is
possible and useful to fix
the relation between this quantity and the light curve. 
That can help us in obtaining  stellar parameters for stars for which
we have only light curves, i.e. the by-product of the microlensing projects.

The Fourier decomposition is a powerful method to describe in a 
synthetic and safe way the shape of a light curve. 
It consists in fitting the photometric measurements by means of the series
\begin{equation}
     I(t)= A_o + \sum_{i=1}^N {A_i \cos [2\pi i(t-T_o)f +\phi_i ]}
\end{equation}
where $I(t)$ is the magnitude observed at time $t$, $A_0$ the mean
magnitude, $A_i$ the amplitude of the $i$--component (the $i-1$ harmonic), 
$f$ the frequency
($f$=1/$P$, where $P$ is the period of the light variation), $\phi_i$ the
phase of the $i$--component at $t=T_o$. 

The Fourier parameters can be subdivided into two groups: the amplitude
ratios $R_{ij}=A_i/A_j$ (i.e. $R_{21}=A_2/A_1$,  $R_{31}=A_3/A_1$,
$R_{32}=A_3/A_2$, ...) and the phase differences $\phi_{ji}=i\phi_j-j\phi_i$
 (i.e. \pdu=$\phi_2-2\phi_1$,
\ptu=$\phi_3-3\phi_1$, $\phi_{32}=2\phi_3-3\phi_2$, ...).
 Note that the use of
the {\tt sin}  term instead of the {\tt cos} leads to results that are not
directly comparable, owing to the $\pi/2$ shift of each component. 
Because it provides quantitative parameters to define the shape of the
light curves, the Fourier decomposition  is a suitable tool for
classification purposes, once care is taken to
avoid spurious results or
misinterpretations of features of  the light curves.
Moreover, the plots of the Fourier parameters versus periods showed the presence of
discontinuities in the progressions, which can be ascribed to
resonances between different pulsational modes, as in the case of 
Classical Cepheids at P$\sim$10~d. These diagrams are therefore useful
tools to perform 
asteroseismology in giant stars, as the Cepheids (Poretti 2000a), even
if some caution is necessary to locate exactly the position of the 
resonances (see Kienzle et al. 1999 and Feuchtinger et al. 2000 for a 
discussion of the resonance
between the fourth and first overtones in the case of $s$--Cepheids).

In this paper, we examine the Fourier plots of the pulsating stars with period 
less than 1~d contained in the OGLE database in order to emphasize their
properties.

\section{The OGLE database}
We considered the 268 pulsating stars included in the 15 fields covered 
by the OGLE database. 
The number of stars is large enough to constitute a conspicuous 
sample, able to provide unambiguous characterizations  of these variables,
and small enough to allow a star-by-star analysis, thus avoiding
the misinterpretations that could arise from automated analysis.  
Each light curve is suitable to be Fourier decomposed
thanks to the high number of homogeneous measurements 
(with a few exceptions) and we expected
a clear detection of features and peculiarities, both for individual stars
and for classes of variables.
The analysis consisted of the  separate steps described in the next
subsections.
\subsection{The verification of previous results}

The primitive idea was simply to investigate the Fourier plots obtained by Morgan et
al. (1998a, 1998b, 1998c) to clarify some ambiguous patterns.
In particular, the \ptu$-P$ plot
does not show any clear progression; the \pdu$-P$ plot also shows a strange 
scatter at short periods.  If these results were confirmed, they
should be considered as a failure of the capabilities offered by the Fourier
decomposition as a sharp tool to describe light curves.

A close examination of the coefficients of the least--squares fit reported
by Morgan et al. (1998a, 1998b, 1998c) reveals that in many cases the small
 amplitude terms are
not significant, the error bars being as large as the amplitudes themselves.
In other cases, the presence of large--amplitude high--order harmonics
(i.e. $A_{i+1}/A_i\sim 1$) implies a  really odd  
fitting light curve  with humps
and bumps. In general, $A_{i+1}/A_i<0.5$ should be the rule far from resonances.  
Only the necessity of fitting the double maximum of RRc variables, 
the well--defined humps at minimum light of some RR Lyr variables or very
asymmetrical curves should involve  deviations
from this rule, generating  a few  consecutive terms having similar 
amplitudes.  Therefore, we matured the idea that a re--analysis of all
the datasets was necessary to handle more reliable Fourier coefficients. At
this point, 
we were forced to scrutinize each star again. 

\subsection{The fit of the OGLE light curves}
\begin{figure}
\resizebox{\hsize}{!}{\includegraphics{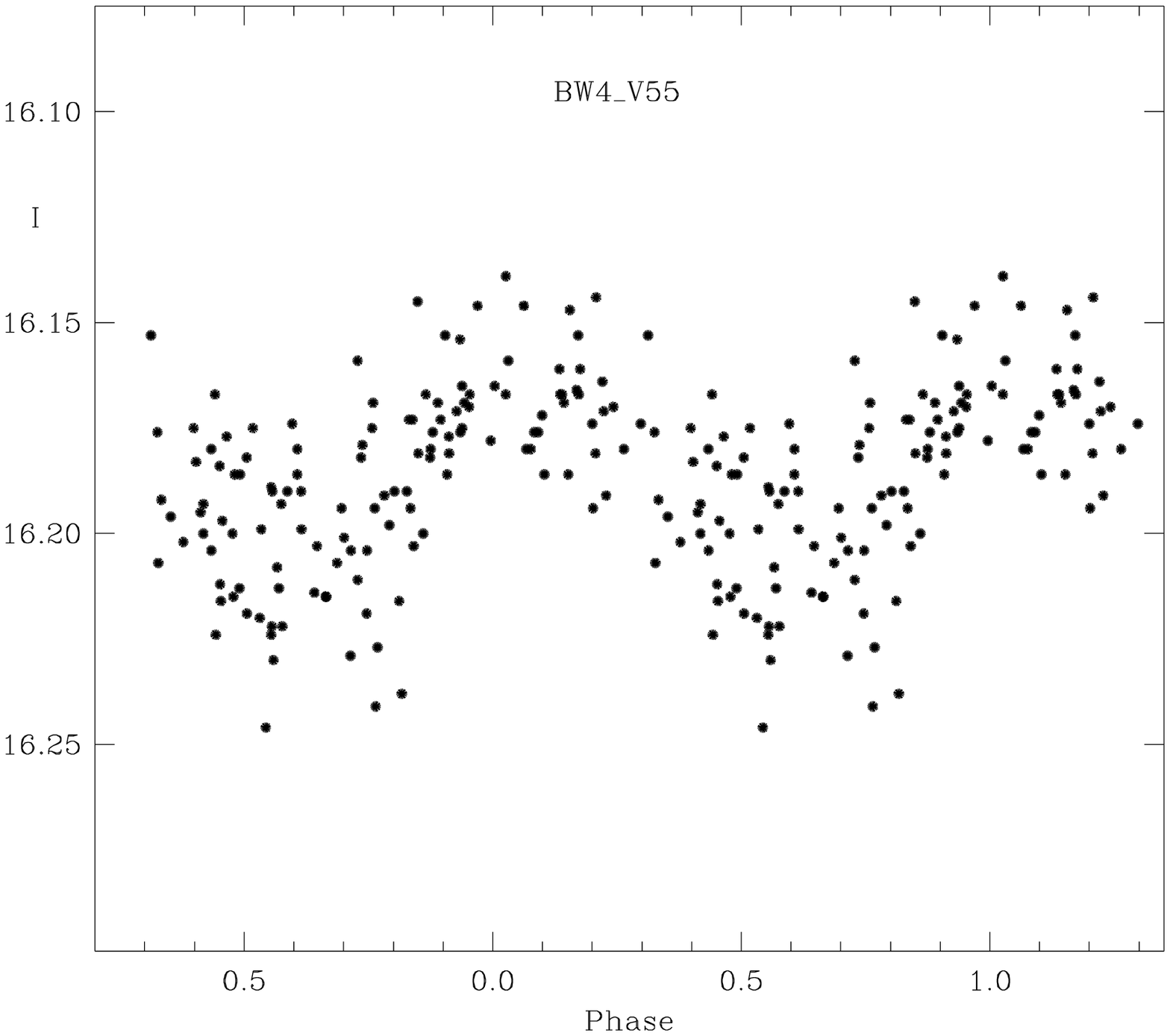}}
\caption[ ]{Example of a sine--shaped light curve (BW4\_V55, $P$=0.15958~d).
The small amplitude (the fitting curve has a full amplitude of 0.04 mag) and the
noise (rms residual 0.016 mag) do not allow the detection of a significant 
first harmonic contribution.}
\label{sinu}
\end{figure}

We excluded 6 stars from the original sample (BW1\_V31, BW9\_V55, BWC\_V22,
BWC\_V28, BWC\_V97 and  BWC\_V56). For these stars,
the number of measurements is less
than 60 and the fit is quite uncertain, hampering a   
reliable determination of the Fourier parameters. 
In 8 other  cases the measurements do not show significant deviations from a
perfect sine--shaped light curve, probably owing to the small amplitude:
BW1\_V90 (full--amplitude 0.08 mag), BW2\_V44 (0.10 mag), BW4\_V55
(0.04 mag), BW4\_V94 (0.06 mag), BW5\_V117 (0.10 mag), BW6\_V120 (0.10 mag),
BW7\_V74 (0.05 mag) and MM5-B\_V141. It should be noted that  0.049~d$<P<$0.180~d 
for all these stars (and for 6 of them $P<0.10$~d): the shorter periods
are compatible with the possibility that these stars are overtone pulsators.
Figure \ref{sinu} shows a light curve in which it was not possible
to detect the amplitude and phase values of the first harmonic in a reliable way. 
Morgan et al. (1998c) considered four of these variables (BW1\_V90, BW4\_V94,
BW6\_V120, BW7\_V74; see their Tab.~1) as unusual $\delta$ Sct.
This misleading
conclusion was due to the over--evaluation of the significance
of their \pdu, \ptu, \pqu parameters.

\begin{figure}
\resizebox{\hsize}{!}{\includegraphics{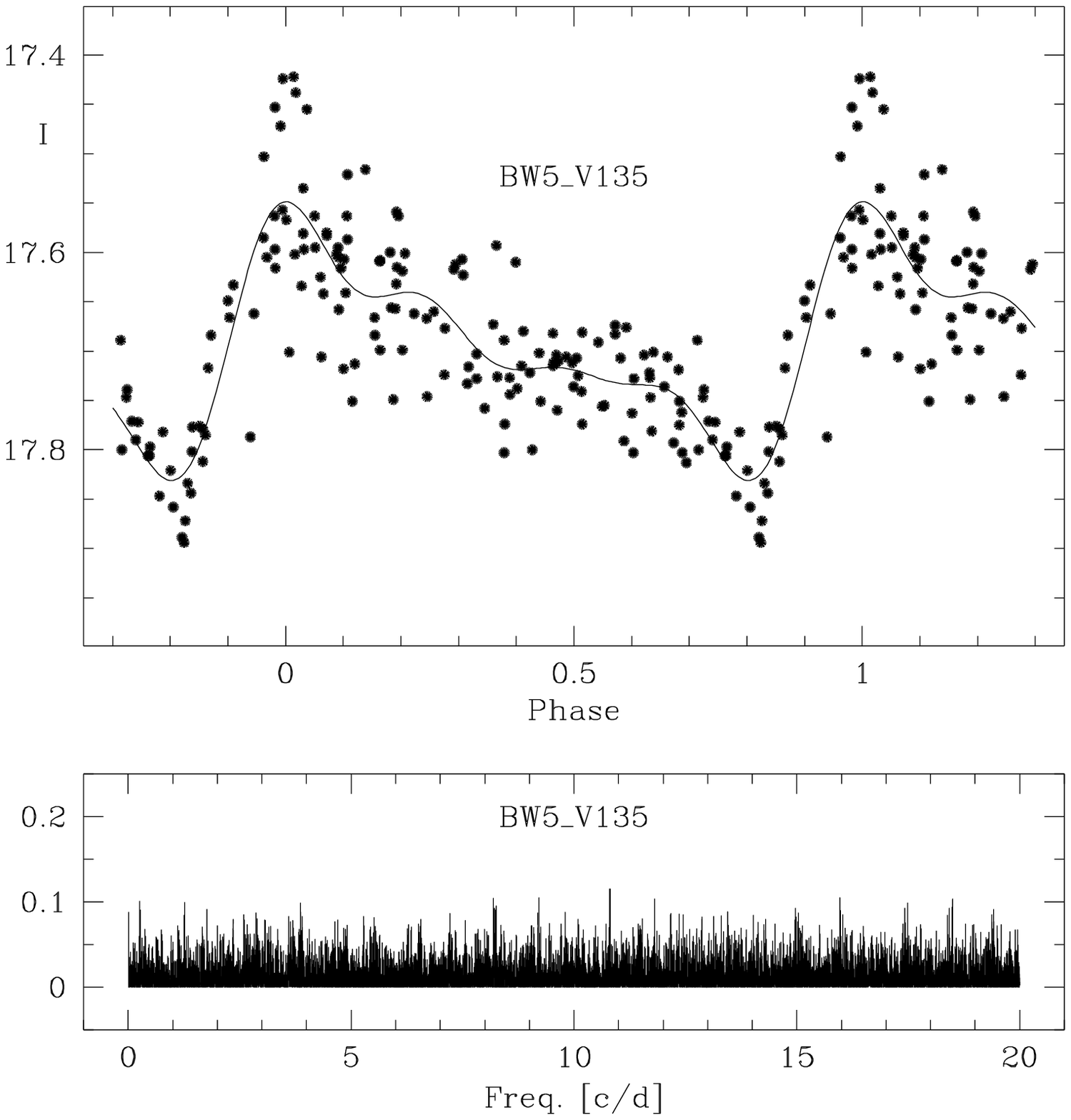}}
\caption[ ]{Top panel: example of an unacceptable least--squares fit
 (BW5\_V135, $P$=0.58825~d).
It is hard to believe that the bumps are real; the asymmetry of the light curve
requires a high order of the Fourier fit ($M$=4) and the 
large scatter enhances the amplitude of the high--order terms.
Bottom panel: the power spectrum of the residuals does not show any significant
second periodicity.}
\label{dege}
\end{figure}

In 8 more cases (BW1\_V7, BW4\_V132, BW5\_V135, BW5\_V174, BW9\_V35,
BW10\_V39, BWC\_V65 and MM7-B\_V7)  it was not possible to apply the 
Fourier decomposition:
a high--order fit was requested to interpolate the measurements
describing a very asymmetric shape, but the result is an improbable curve with 
evident bumps.  In such cases the combination of the noise with the large number
of free parameters is responsible for the unacceptable (from an astrophysical point
of view) curve, even if the least--squares routine brings out figures.  
 The frequency analysis did not reveal any second periodicity. Figure~\ref{dege}
shows an example.

Finally, we reduced to 246 the number of stars whose light curves were 
able to be decomposed.
The 22 stars we rejected can be considered
as partially responsible for the unclear Fourier plots obtained by Morgan
et al. (1998a, 1998b, 1998c). In particular, this led us to a first specific result.
We had already found that 4 of the stars described by Morgan et al. (1998c) as
unusual $\delta$ Sct stars have a sine--shaped light curve. Proceeding further, 
we found  that the curves of all the remaining ``unusual" stars 
could be fitted by a $M$=2 Fourier series and
hence the \ptu and \pqu values  don't have any significance. Since the \pdu
parameters of these remaining stars assume quite normal values, the claimed
subclass of 
unusual $\delta$ Sct stars previously reported doesn't exist.
 
Moreover, we changed the decomposition order $M$  of about 70\% of stars;
in this way the Fourier plots involving high--order terms improved in clearness. 
\subsection{The period refinement}

The Fourier decomposition of the light curves by a single frequency (plus
harmonics) could not always give a satisfactory fit, sometimes leaving a 
high rms residual. In such cases, it is quite obvious to investigate the
datasets searching for other periodicities; this examination had not been 
done in the previous analysis. In many cases our frequency analysis revealed
the presence of other peaks in the power spectrum. 
Since most of  these peaks are observed very close to the main one  and
since such periods are expected as signatures  both of
nonradial oscillations and of a Blazhko effect, we must be sure that the reported
periods are accurate. If not, a wrong period can generate spurious alias peaks
or change the observed value of the neighbouring real peaks. 

 Therefore, we decided to refine
the periods of all the stars and then to re-analyze in frequency all the datasets. 
To this end, we considered as preliminary solutions  the least--squares
fit obtained by using the period values reported in the OGLE database
and the order $M$ established in the previous step.
This solution was given as input parameters to a code keeping
locked the relations between the main frequency and the harmonic terms
(MTRAP; Carpino et al.  1987);  
the best fit was searched for around the preliminary solution. 
Once the correct period was obtained, we re--examined in frequency all the
datasets to detect secondary periodicities. 
No prewhitening was done: only the frequency values of the main term and its
harmonics were considered as input values (known constituents), not their
amplitude and phase. This approach 
is the same as the one used by Pardo \& Poretti (1997) to detect the frequency
content of double--mode Cepheids. The period refinement was marginal in
most cases: for 73 stars (30\% of the sample) it is smaller than $1\cdot10^{-6}$~d,
for 168 stars
(72\% of the sample) is smaller than $5\cdot10^{-6}$~d. The mean error on
the period is $1.3\cdot10^{-6}$~d and for about 100 stars out of 234 the refinement was made
within this error bar. As a by-product, we made  a further revision of the 
significance of the small amplitude terms.
The main result of this search is described in
the next subsection. 
 
\subsection{The different frequency content of pulsating variables}
\begin{figure}
\resizebox{\hsize}{!}{\includegraphics{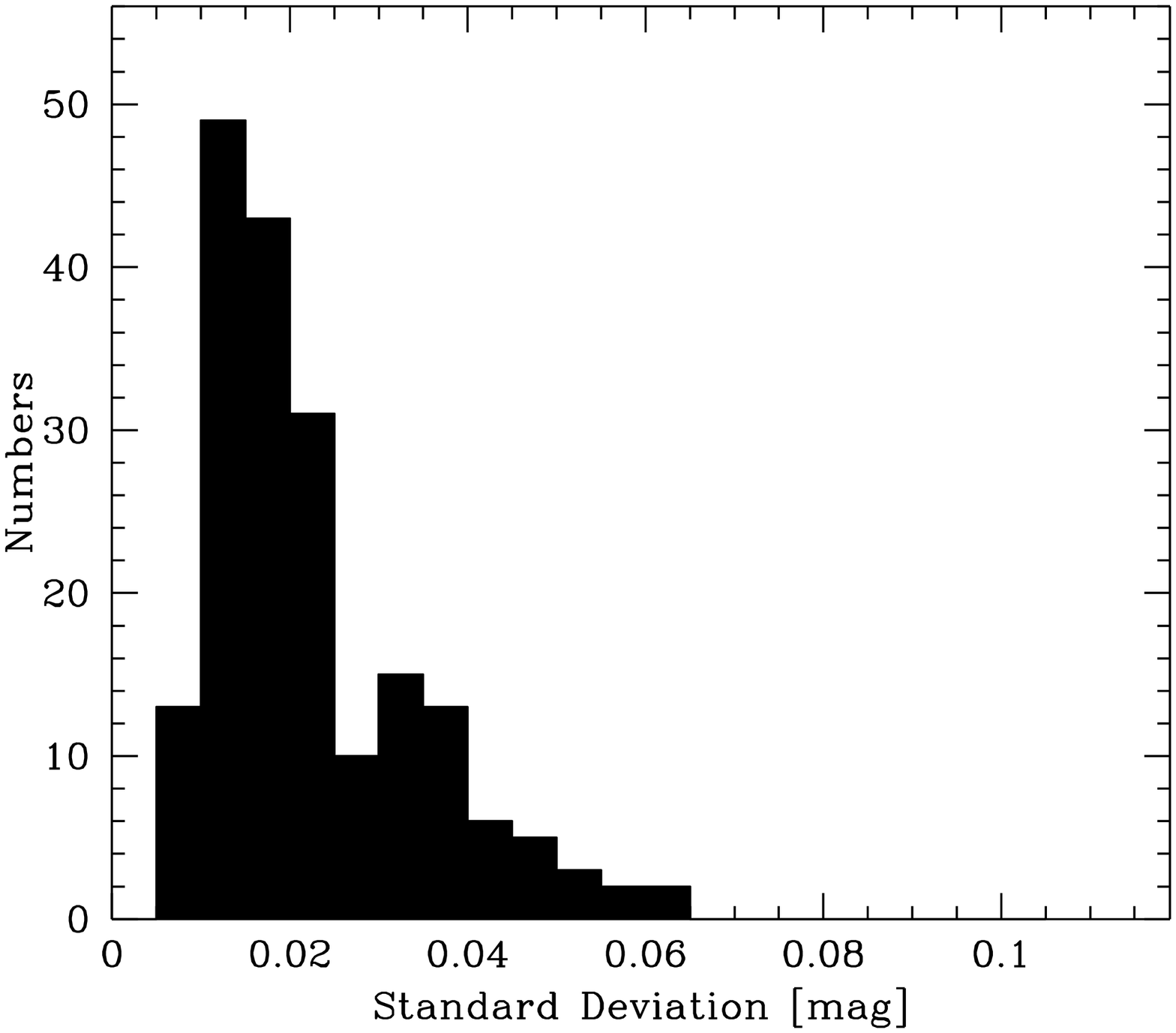}}
\caption[ ]{ Histogram of the standard deviations of the reliable Fourier
fits of the light curves of the 192 monoperiodic stars.}
\label{sd}
\end{figure}

At the end of this step we separated the monoperiodic variables from the
other stars.  Of course, a second,
small-amplitude term may always be hidden in the noise, even if the fit 
with a single--period looks satisfactory. However, after the further check
we made, we can be confident that its effect on the main period should be negligible.
As a result of the  period refinement procedure and the subsequent frequency analysis,
we evidenced 53 stars showing a well--established multimode behaviour:
\begin{enumerate}
\item double--mode stars pulsating in the fundamental radial mode
   and in another  radial overtone (7 candidates), with a well--separated 
   pair of frequencies;
\item stars showing one single peak close to the main frequency (30 
   candidates), with a
   ratio \fu/\fd$\sim$0.99 or $\sim$1.01;
\item stars showing two peaks symmetric with respect to the main
   frequency and, in some cases, to the harmonics (7 candidates);
\item stars showing a second peak
   very close to the main one (9 candidates), \fu/\fd$\sim$0.999 or 
   $\sim$1.001, suggesting
   a period variation or an unresolved, very long Blazhko effect.
\end{enumerate}
All these types of multimode variables will be discussed in forthcoming papers.
Not considering the Pop.~II Cepheid BWC\_V1 ($P$=1.74797~d), we got a
reliable Fourier decomposition of 192 monoperiodic variables with $P<$1~d.  
In general, the fit is satisfactory even if some light curves appeared to
be noisy. That can be ascribed both to difficult image reduction (close
companion, nearby bright stars,~...) or to slow instrumental
drifts, which appear in the frequency analysis as peaks very close to
0.000~\cds or 1.000~\cd. In 136 cases 
(i.e. the 71\%) the least--squares fits have a standard deviation better 
than 0.03 mag (Fig.~\ref{sd}),
a value quite acceptable for stars of $I$--magnitude 15--17 
measured on a time baseline of three years. 

We would like to emphasize the fact that all the work described above constitutes
a complete independent re--analysis of the whole OGLE database. The 
new  multiperiodic pulsators we found, the sharper revision of the light curve
fits and the elimination of unclear subclasses of variables 
configure our Fourier analysis of the OGLE database as a very reliable
tool to investigate the properties of the pulsating variables with $P<1$~d.

\section{The Fourier diagrams: the separations between variable classes}

We can finally handle the well--defined Fourier plots
shown in Figs.~\ref{f21}, \ref{f31}, \ref{f41} and \ref{r21}.
Table~\ref{erro} lists the averaged errors on the Fourier parameters
for each class, as calculated from the errors on the least--squares
fits.
\begin{center}
\begin{table}
\caption{Averaged errors on the \pdu, \ptu, \pqu~and \Rdu~ parameters.}

\begin{flushleft}
\begin{tabular}{l r r r r} 
\hline
\multicolumn{1}{c}{}&\multicolumn{1}{c}{\pdu}&
\multicolumn{1}{c}{\ptu}&\multicolumn{1}{c}{\pqu}&
\multicolumn{1}{c}{\Rdu}\\
\cline{2-4}
\multicolumn{1}{c}{}&\multicolumn{3}{c}{[rad]}&
\multicolumn{1}{c}{}\\
\hline
\noalign{\smallskip}
HADS        &   0.23   & 0.34 &  0.41 & 0.05 \\
RRc         &   0.16   & 0.24 &  0.33 & 0.02 \\
RRab        &   0.05   & 0.07 &  0.12 & 0.02 \\
\hline
\end{tabular}
\end{flushleft}
\label{erro}
\end{table}
\end{center}

Considering $P<$1 d, the pulsating stars should be subdivided into
High--Amplitude $\delta$ Scuti (HADS), RRc and RRab stars. However,
the continuity of the progressions of the phase parameters
(Figs.~\ref{f21}, \ref{f31}, \ref{f41}) masks the separation 
between HADS and RRc stars. Such a subdivision is more appreciable in the
\Rdu$-P$ plot (Fig.\ref{r21}).
 As  a first step, a close examination
can help in separating variable classes more precisely.
For the sake of clarity, the subdivisions are shown right away in 
Figs.~\ref{f21}, \ref{f31}, \ref{f41} and \ref{r21} by means of
different symbols.

\subsection{HADS or RRc ?}
\begin{figure}[p]
\includegraphics[width=6.8truecm]{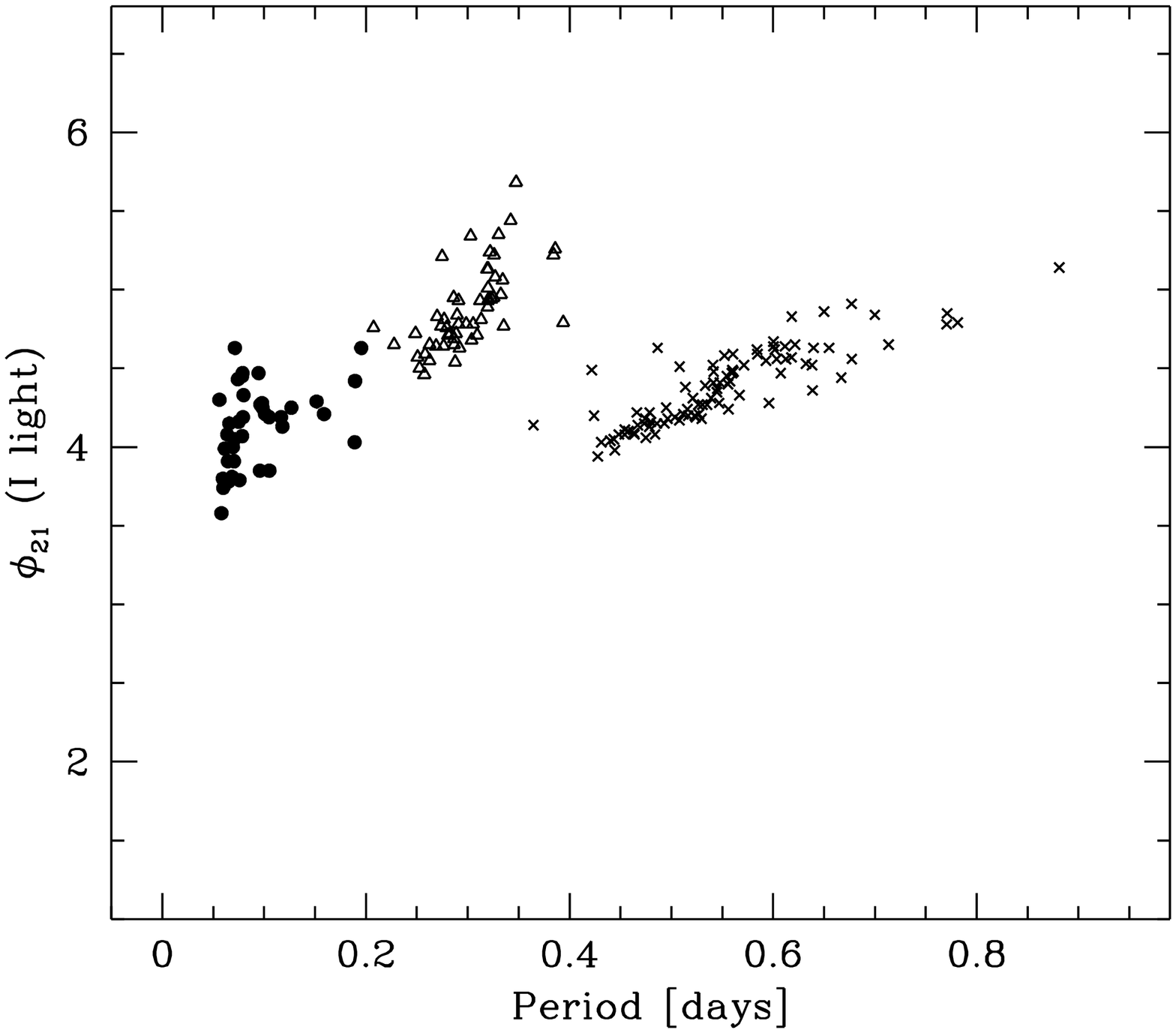}
\caption[ ]{Fourier parameter \pdu against Period for the 192
monoperiodic stars with $P<$1~d in the OGLE database. Filled circles:
HADS; open triangles: RRc; crosses: RRab stars.}
\label{f21}
\end{figure}

\begin{figure}[p]
\includegraphics[width=6.8truecm]{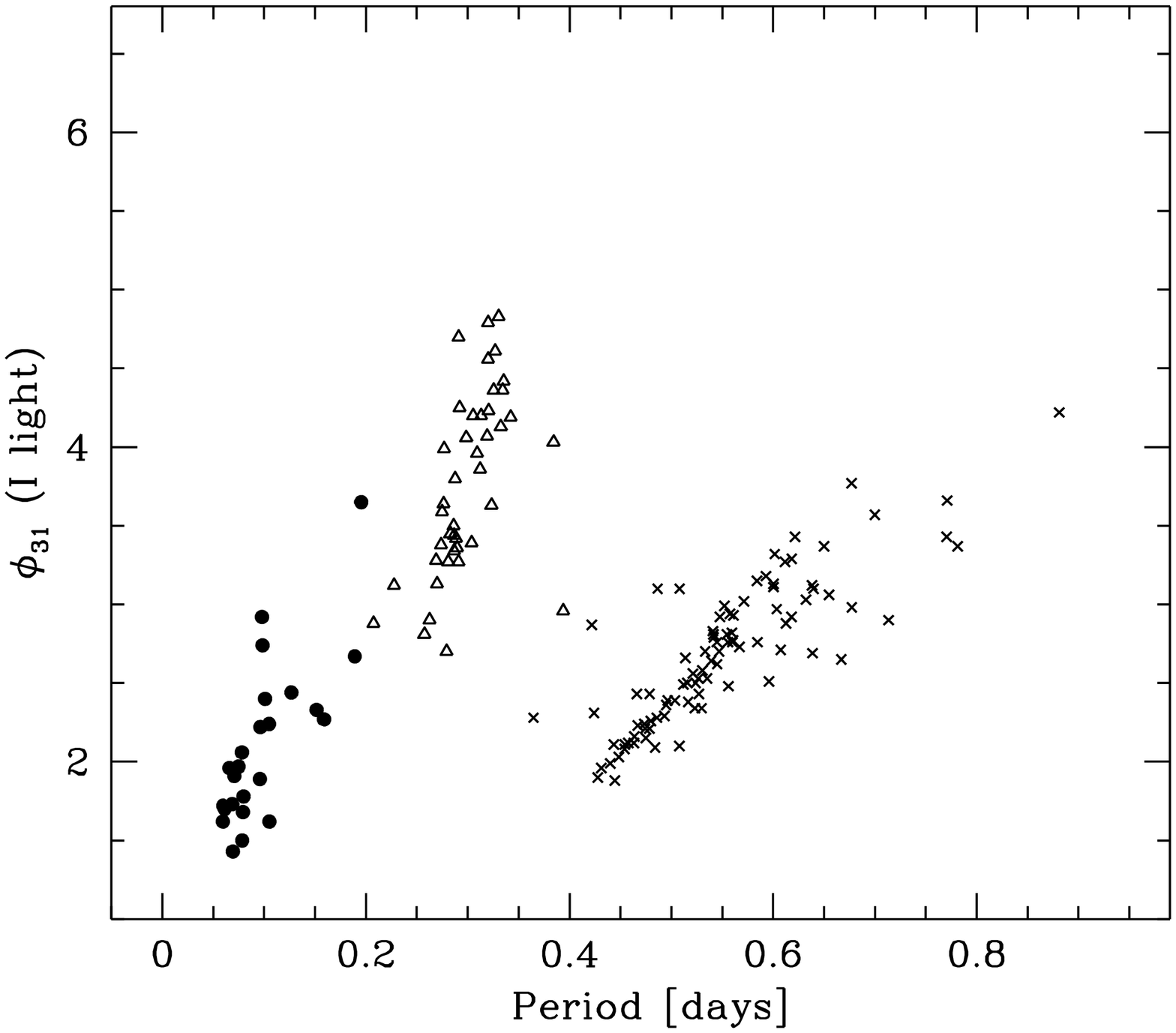}
\caption[ ]{Fourier parameter \ptu against Period for the 192
monoperiodic stars with $P<$1~d in the OGLE database. Same symbols as
in Fig.~\ref{f21}.}
\label{f31}
\end{figure}

\begin{figure}[p]
\includegraphics[width=6.8truecm]{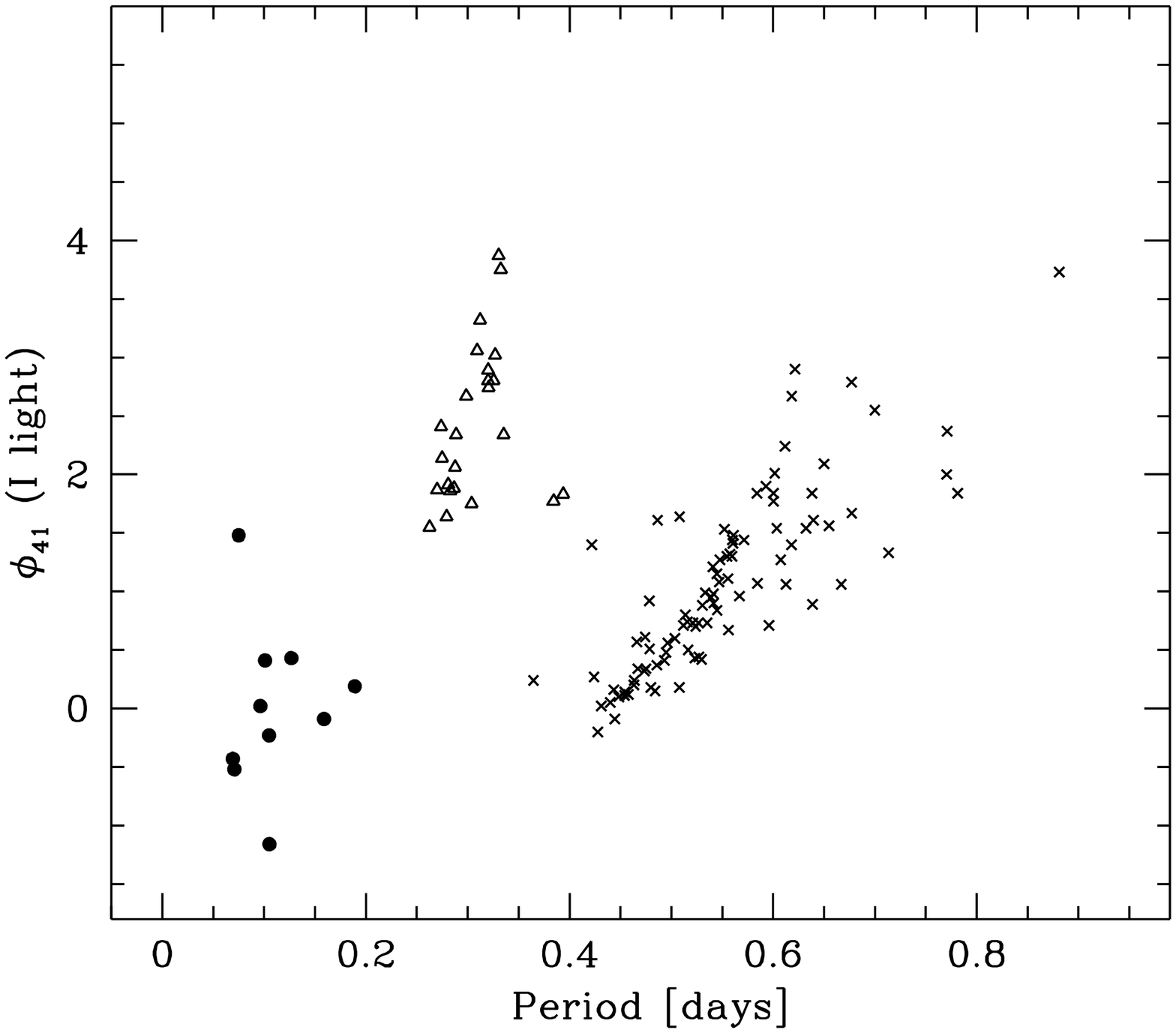}
\caption[ ]{Fourier parameter \pqu against Period for the 192
monoperiodic stars with $P<$1~d in the OGLE database. Same symbols as
in Fig.~\ref{f21}.}
\label{f41}
\end{figure}

\begin{figure}[p]
\includegraphics[width=6.8truecm]{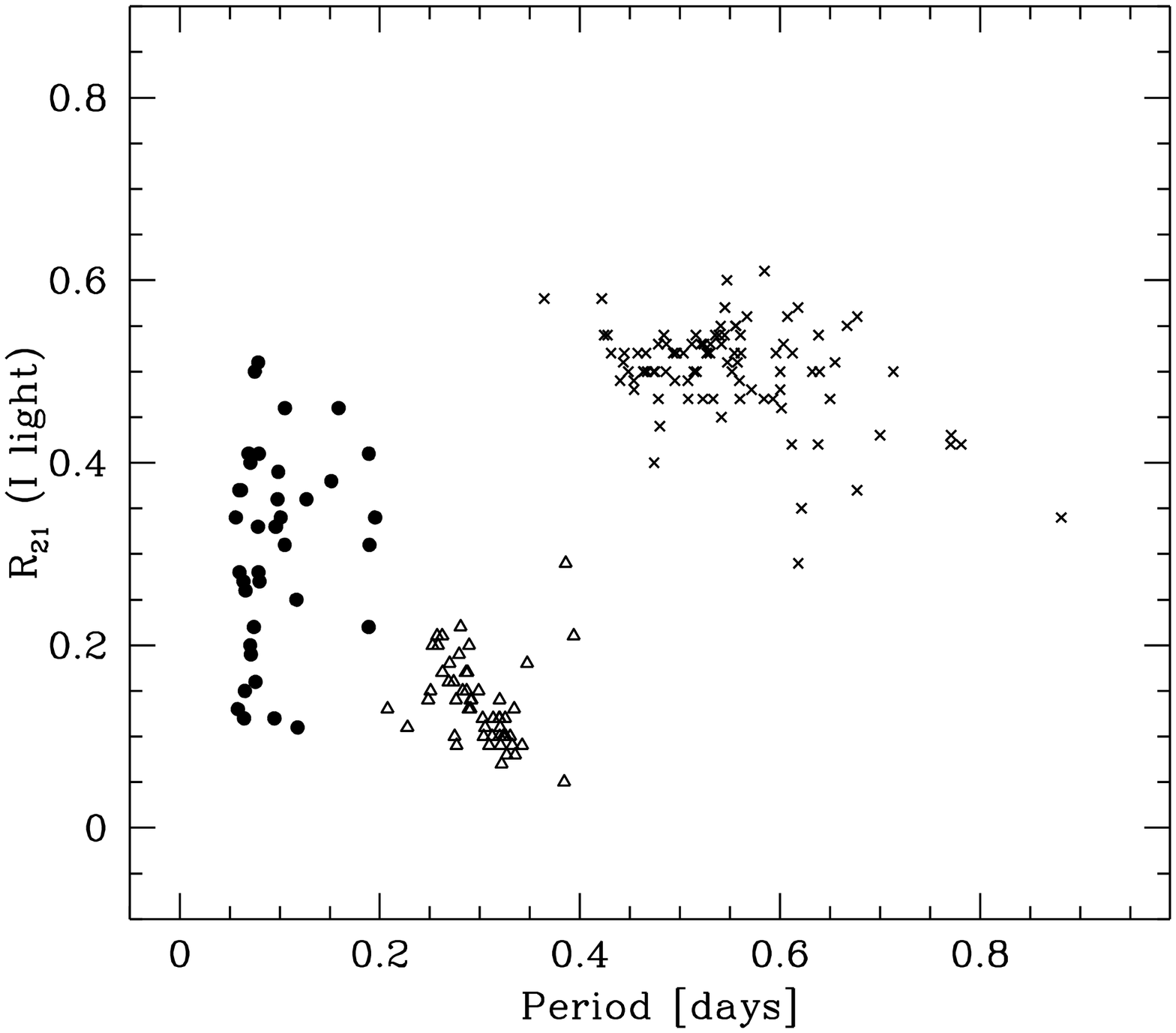}
\caption[ ]{Fourier parameter \Rdu against Period for the 192
monoperiodic stars with $P<$1~d in the OGLE database. This plot yields the
clearest separation between HADS and RRc stars. Same symbols as
in Fig.~\ref{f21}.}
\label{r21}
\end{figure}

\begin{figure}
\resizebox{\hsize}{!}{\includegraphics{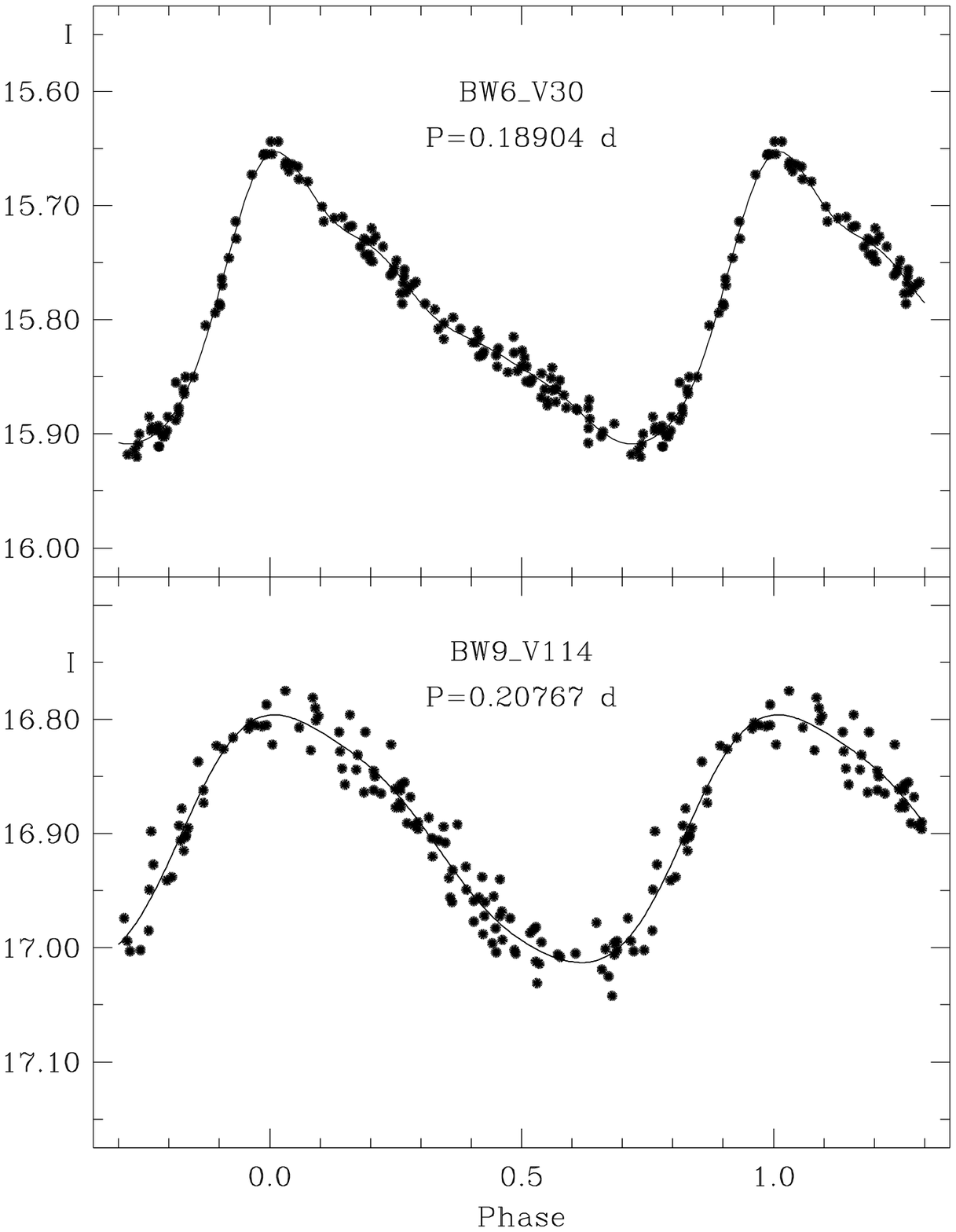}}
\caption[ ]{Top panel: note the sharp maximum in the light curve of
BW6\_V30 ($P$=0.18904~d), which can be considered the HADS with the longest
period.  Bottom panel: note the rounded maximum in the light curve of BW9\_V114
($P$=0.20767~d), which seems to be the extreme short--period RRc.}
\label{harr}
\end{figure}

In the period range 0.15-0.25~d, separating an RRc variable from a HADS is
not an easy task, since for example, the \pdu$-P$ plot (Fig.~\ref{f21})
shows a merging of two families of points, without any apparent
jump. However, the light curve of a star following the progression coming from the
shortest period shows  an asymmetrical light curve with a sharp maximum
 (BW6\_V30, $P$=0.18902~d; Fig.~\ref{harr}, top
panel), while in the opposite direction we
generally find more smoothed light curves with rounded maxima (BW9\_V114,
$P$=0.20767~d, Fig.~\ref{harr}, bottom panel). If the difference is not
quite evident in the \pdu$-P$ and \ptu$-P$ plots (Figs.~\ref{f21} and
\ref{f31}), it can be better
appreciated both in the \pqu$-P$ plot (Fig.~\ref{f41})
and in the \Rdu$-P$ one (Fig.~\ref{r21}). 

We didn't detect any $A_4$ term for stars in the 0.20~d$<P<$0.25~d
interval, suggesting that the light curves are more sinusoidal: 
it looks natural to ascribe this feature to the rapid evolution
of the light curves of RRc stars (see the steady slope of \pqu parameters for
$P>0.30$~d). 
In the latter, a well defined dip is  observed in the interval from 0.20 to
0.40~d, suggesting that all these stars belong to the same family. In
particular, note the \Rdu$>$0.2 values for 
$P<$0.20~d and the \Rdu$<$0.15 for 0.20~d$<P<$0.25~d. We can conclude that
the Fourier parameters of stars  in the interval 0.20~d$<P<$0.25~d are
naturally linked with the longer periods (i.e. RRc stars) rather than with
the shorter ones (i.e. HADS stars).

\clearpage

\begin{figure}
\resizebox{\hsize}{!}{\includegraphics{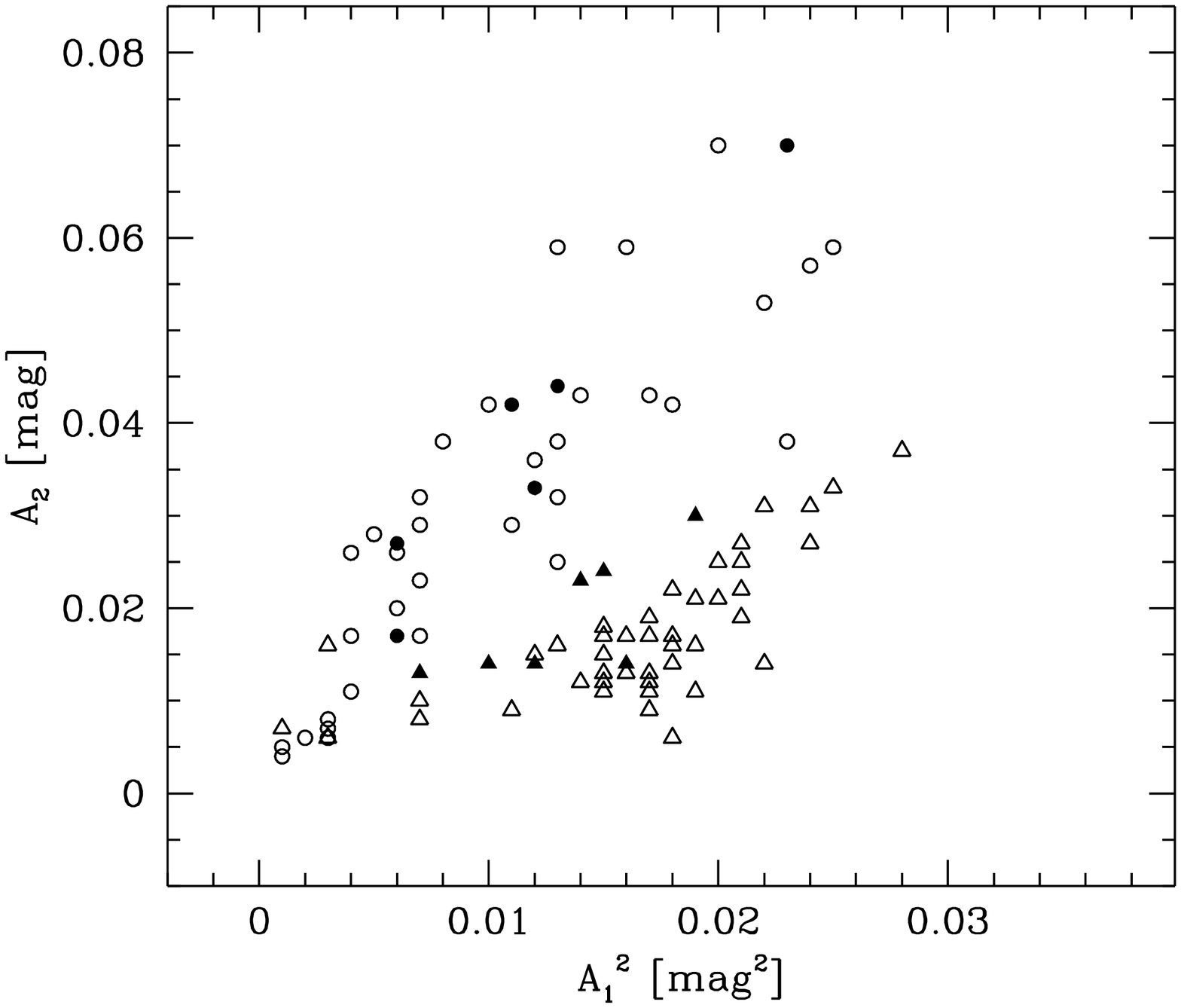}}
\caption[ ]{The amplitudes of the first Fourier terms separate HADS
(circles) from RRc (triangles) stars. Filled circles are HADS stars with
0.15$<P<0.20$~d; filled triangles are RRc stars with 0.20$<P<0.25$~d.}
\label{amp}
\end{figure}

Another way to separate HADS from RRc stars is to plot the $A_1^2$ term versus
the $A_2$ one (Fig.~\ref{amp}, see also Antonello 2000). 
There is some crowding close to the origin, but the two classes are well separated
in the critical range 0.15--0.25~d.

Therefore, we have demonstrated that when dealing with a well--defined sample
of accurate light curves, it is possible to separate
HADS from RRc stars not on the basis of a mere assumption (Antonello 2000),  but by  
looking at the amplitude values. We shall find  below a further
confirmation of this separation. 

The separation between HADS and RRc stars by using $P$=0.20~d as a clean cut
cannot be considered as an inviolable rule; as an example, the symmetrical
light curve of MM5-B\_V141 ($P$=0.181~d) suggests that this star belongs to
RRc variables rather than to HADS. 
We also note that the presence of  second overtone
pulsators among RRc variables is always an open possibility; such pulsators
should show  more
symmetrical light curves and the low $R_{21}$ values observed here can be
a hint. 

\subsection{RRc or RRab ?}
Looking at Fig.~\ref{f21}, RRc and RRab stars are clearly separated. In
general, RRc stars have a period $P<$0.35~d and for this period
\pdu$>$4.5 rad; RRab stars have a period $P>$0.42~d and 
for this period \pdu$<$4.0 rad. Similar separations can be found in 
all the phase and amplitude plots. 

\begin{figure}
\resizebox{\hsize}{!}{\includegraphics{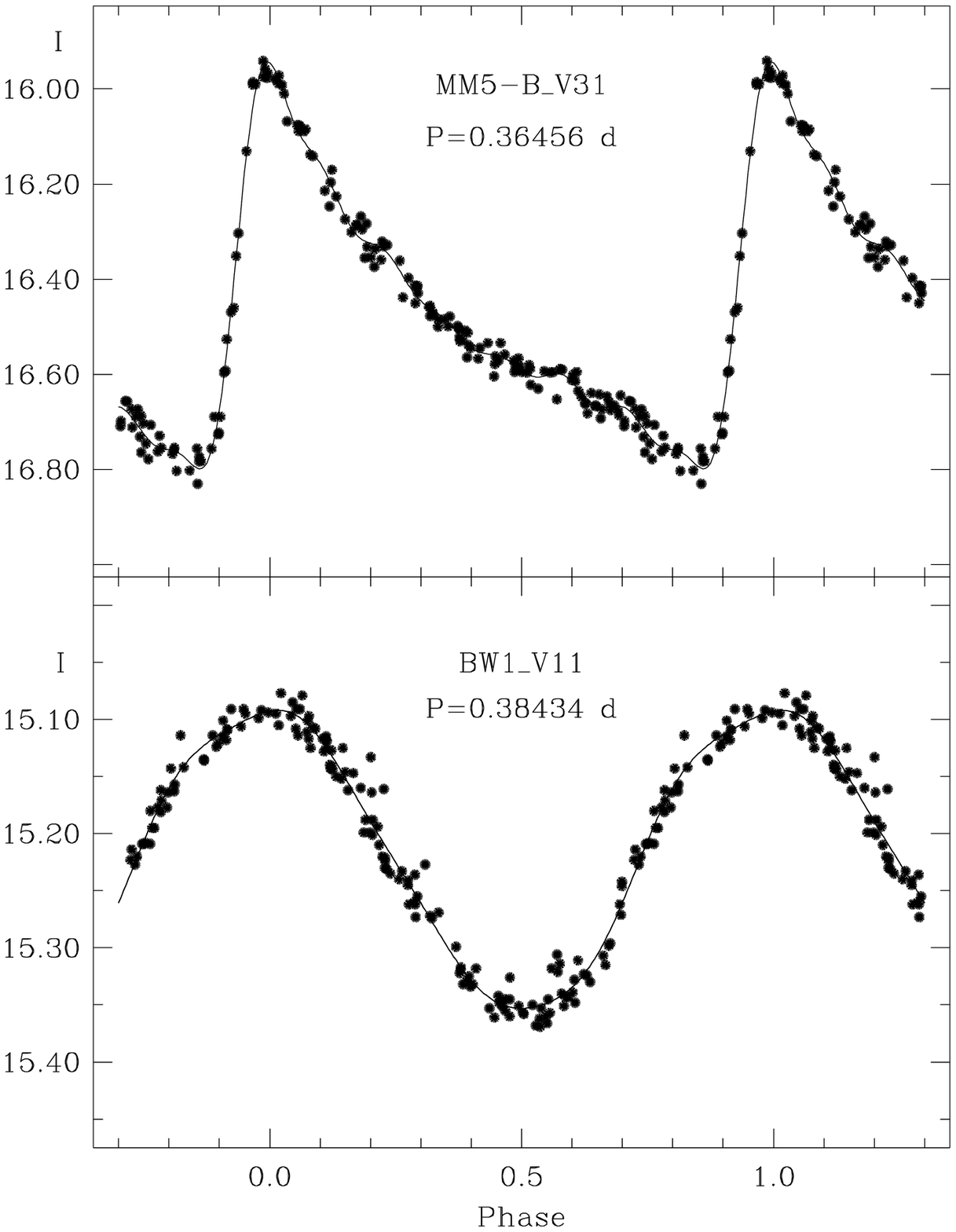}}
\caption[ ]{Top panel: MM5-B\_V31 is the RRab star having the shortest
period ($P$=0.36456~d).  
Bottom panel: BW1\_V11 is the RRc star having the longer period ($P$=0.38434~d).
Note also the strong difference in amplitude.}
\label{bca}
\end{figure}

There are only a few exceptions, i.e. the presence of an isolated
short--period RRab star, MM5-B\_V31 ($P$=0.36456 d) and of three long--period
RRc stars, BW1\_V11 ($P$=0.38434~d), BWC\_V81 ($P$=0.38588 d) and BW2\_V8
($P$=0.39402~d). 
However, the differences in the light curves are quite evident also in these
intermediate cases (Fig.~\ref{bca}). The $R_{21}-P$ plot provides the best
separation between short--period RRab stars and long--period RRc stars 
(Fig.~\ref{r21}).
While the RRab star displays normal values for its class,
the three RRc stars display intermediate values, but always 
closer to RRc-- than to the RRab--type (especially in
the \Rdu$-P$ plot). It should be noted that there are few stars in the
period interval from 0.35~d to 0.40~d; therefore it is probable that
these stars have some particularity.

\section{The Fourier diagrams: insights into the different classes}
The large  number of stars forming each class allows us to discuss some
aspects in detail. 

\subsection{HADS stars}

The \pdu values are quite constant from 0.07~d
to 0.20~d, but they show a slight trend to decrease toward shorter periods.
For the 11 stars with $P<$0.07~d, we get \pdu=3.92$\pm$0.06 rad,
while for the 28 stars with 0.07~d$<P<$0.20~d we get \pdu=4.23$\pm$0.04. This
tendency was already noticed in the $\omega$ Cen stars (Poretti 1999), where 
the presence of stars with shorter period (down to 0.036~d) makes the trend
more evident.
When considering the variables with $P<$0.05~d, the slope of the \pdu$-P$
progression
is not only confirmed, but seems to become steeper again, suggesting a second
break (see also Fig.~1 in Poretti 2000b).
The change in the slope can  also be seen in the \ptu$-P$
and \pqu$-P$ plots, but it is difficult to place it. The gap in the period between
0.080~d and 0.094~d is a further complication. It should be noted that such a gap
is not filled by considering variables we excluded for some reason or other: 
the only star in this interval is the double--mode pulsator BW1\_V207.

In correspondence with the break in the \pdu$-P$ progression, Antonello (2000) 
noted a saddle (centered at $P$=0.10~d and
flanked by two maxima at 0.08~d and 0.14~d) in the amplitude plots. He
correlated this feature  with
the stage of stellar contraction before the expansion of the envelope. As a matter
of fact, the  change of slope is unlikely to be due to a resonance effect or
a different pulsation mode, which both should produce much more evident changes in the
light curves. As regards the latter possibility, we remind the reader of the
presence of perfectly sine--shaped light curves in this period range, which could be the
signature of an overtone radial mode.

\begin{figure}
\resizebox{\hsize}{!}{\includegraphics{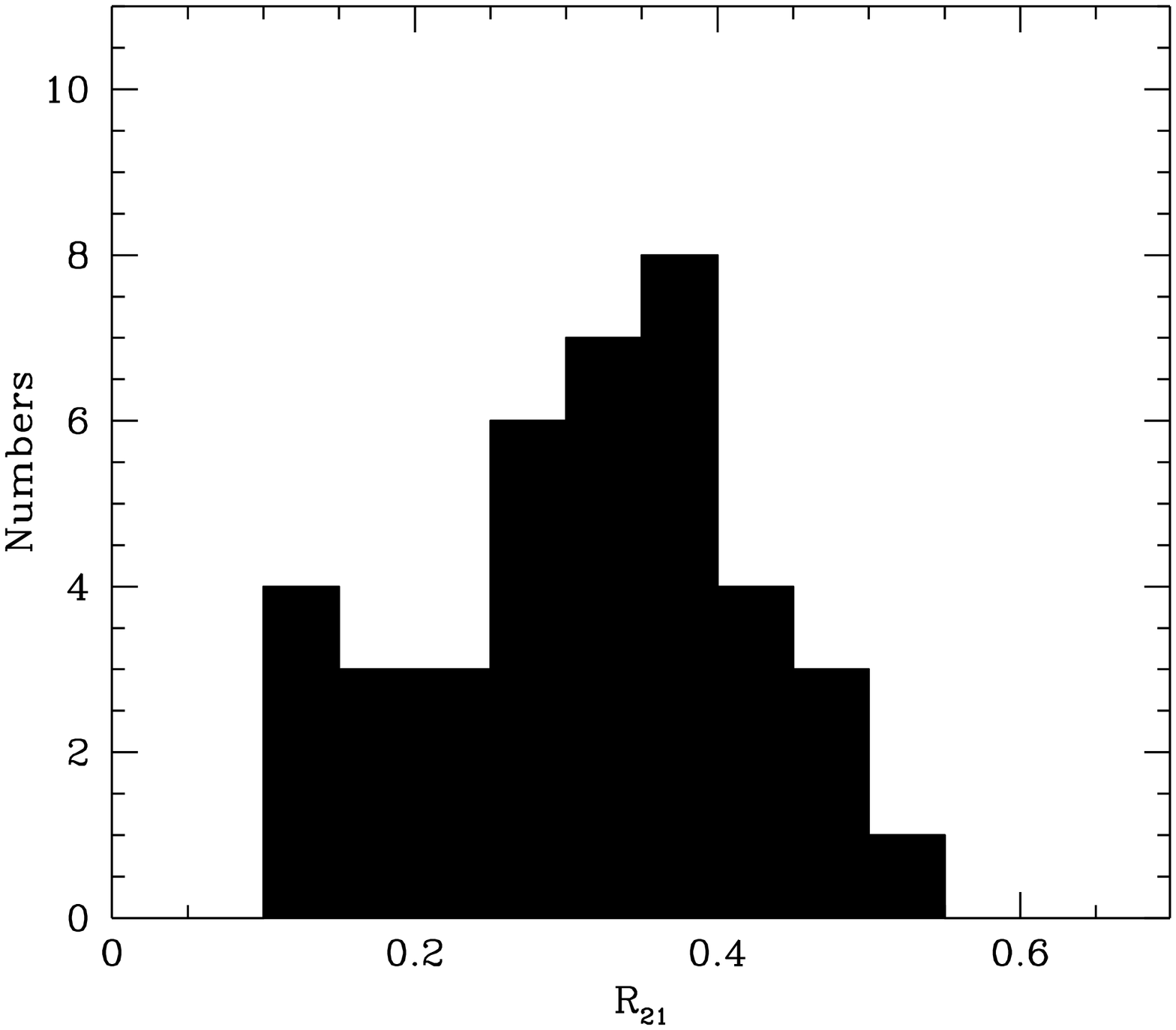}}
\caption[ ]{Histogram of the \Rdu values observed in the light curves of HADS stars.
The distribution does not show any peculiarity considering stars with 
0.20~d$<P<$0.30~d as RRc stars.}
\label{hist}
\end{figure}

Another feature of interest is the bimodal distribution of the \Rdu values
found by Antonello et al. (1986), questioned by Hintz \& Joner (1997), re--confirmed
by Musazzi et al. (1999), simultaneously weakened by Morgan et al. (1998c) and
strengthened by
Templeton et al. (1998), to be finally found unclear by Templeton (2000). 
When plotting the values obtained from the OGLE
stars with $P<$0.30~d (as in the Antonello et al. paper), we observe a bimodal
distribution (with different maxima). However, when considering the stars with
$P<$0.20~d (i.e. the genuine HADS stars, as previously established), the distribution
loses that characteristic, showing a single maximum
toward high \Rdu\, values (Fig.~\ref{hist}). 
Such high values are required to explain the strongly asymmetric light curves of
HADS stars. Therefore, the bimodal distribution is generated by the low \Rdu values 
typical for RRc stars having short periods, as seen above. This is also
confirmed by looking at
the Tab.~2 reported by Antonello et al. (1986): the \Rdu$<$0.20 values  are mostly
found in stars with $P>$0.20~d.  As an interesting feedback,
this results strengthens our
confidence in considering \Rdu=0.20 as the separation between HADS and RRc for
periods around 0.20~d. 

\subsection{RRc stars}
 There are no peculiarities in the light curves of RRc stars.
The only exceptions are the short-- and long--period stars which are
respectively close to HADS and RRab variables and whose parameters 
show intermediate values in some cases.  The Fourier parameters are
well confined in narrow loci in all the plots (Figs.~\ref{f21}, \ref{f31}, \ref{f41}
and \ref{r21}) and their progressions describe very well the main characteristic 
of these stars, i.e the slow but continuous change in the shape of the light maximum, 
which appears to be double or flat in most cases. When the light curves  are
more rounded, i.e. around 0.31--0.32~d, there is a corresponding dip
in the $R_{31}-P$ and $R_{41}-P$ diagrams. 

\subsection{RRab stars}
\begin{figure}
\includegraphics[width=7.0truecm]{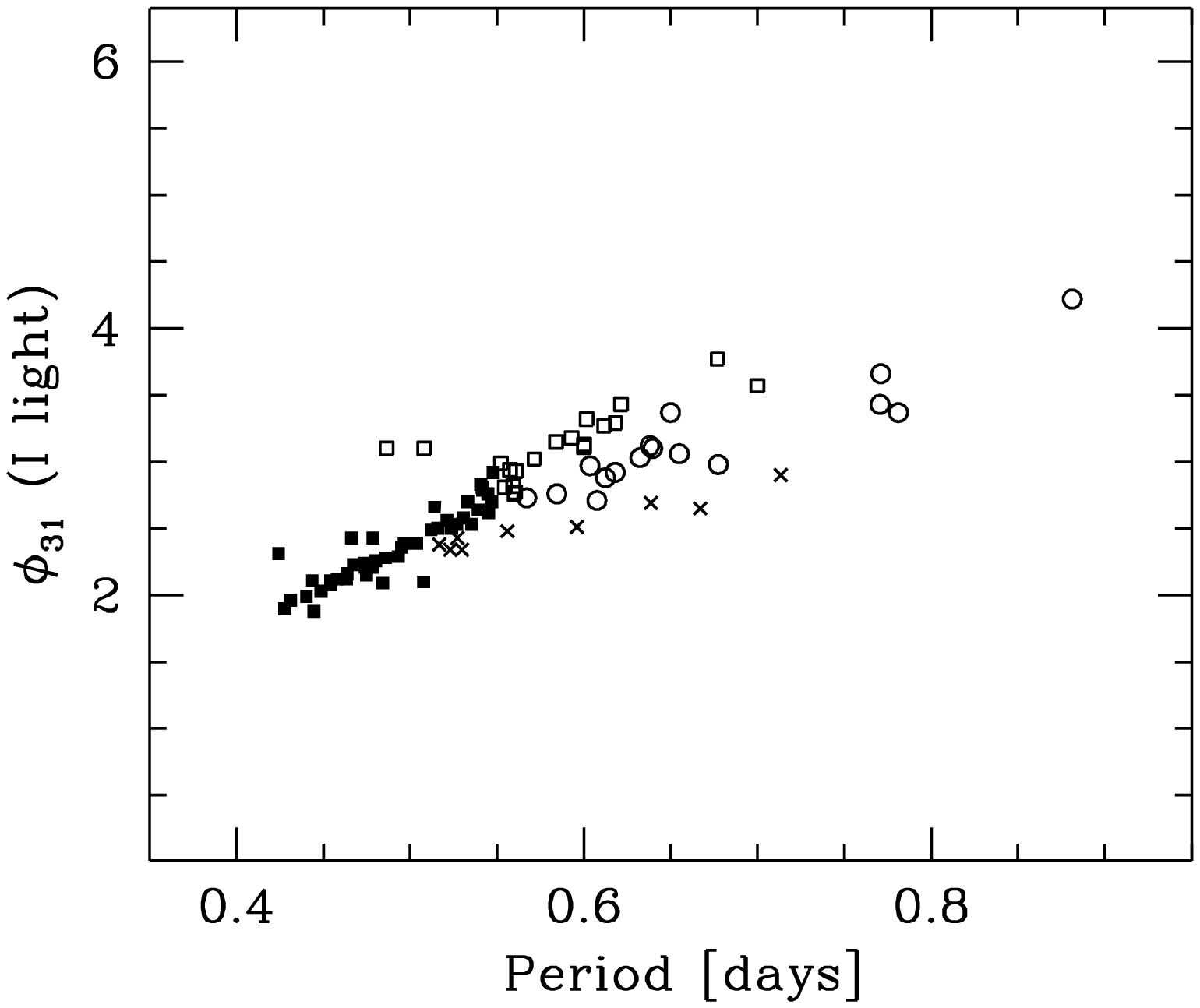}
\includegraphics[width=7.0truecm]{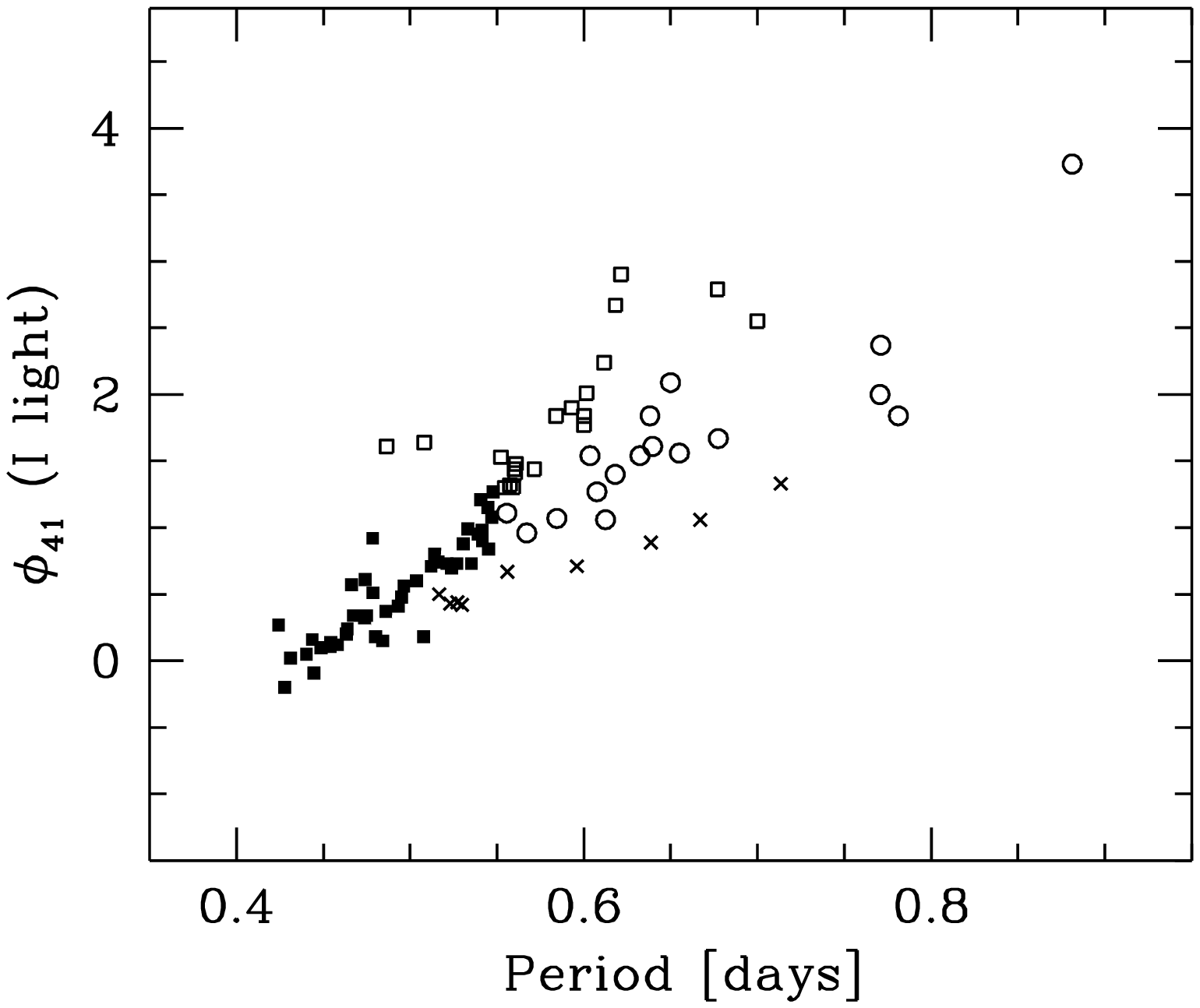}
\includegraphics[width=7.0truecm]{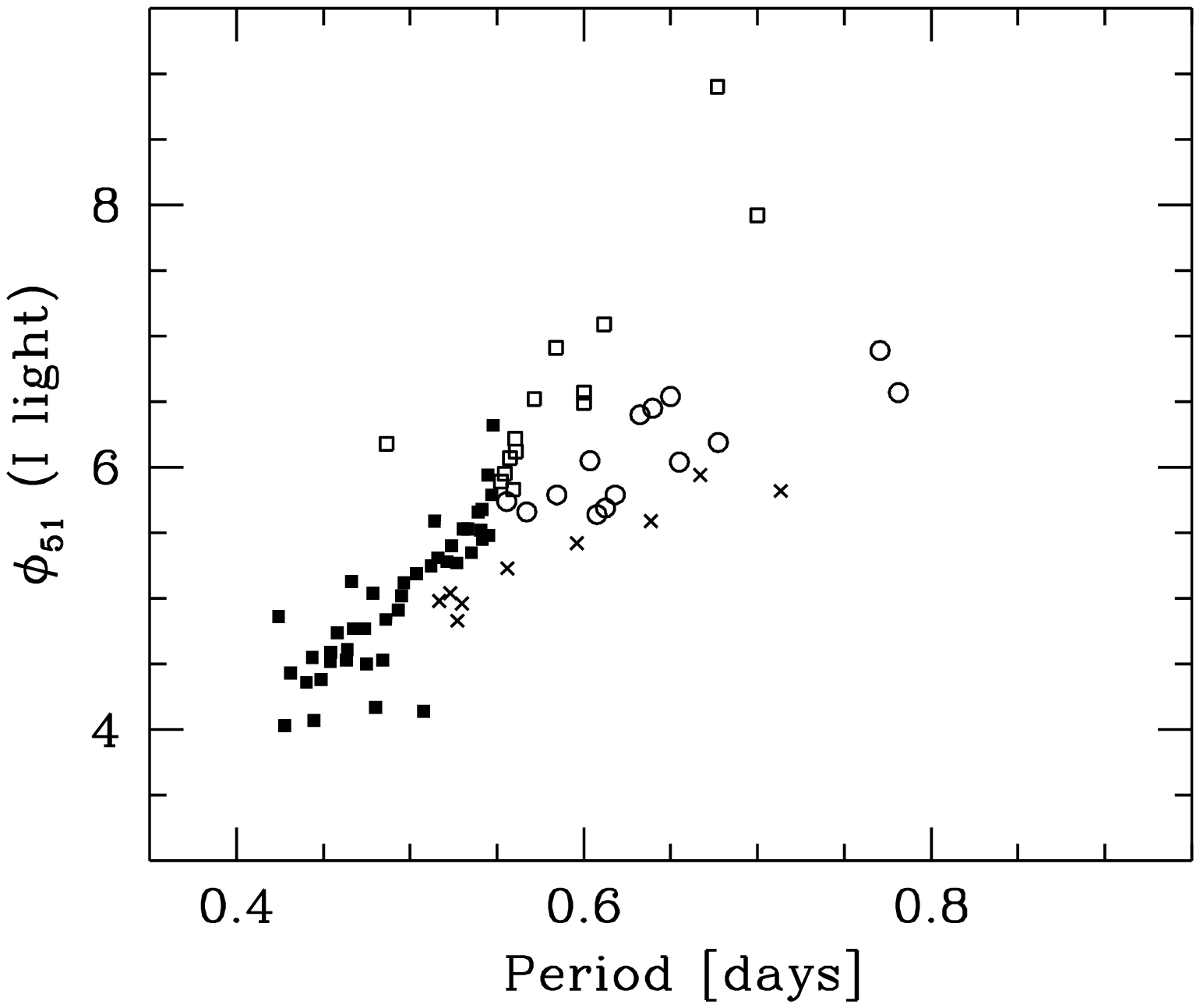}
\caption[ ]{Fourier \ptu, \pqu and $\phi_{51}$ parameters of 
RRab stars in the OGLE database ($I$ light). After a common
progression until $P$=0.50--0.55~d (filled squares), the $\phi_{j1}$ 
values diverge into different ``tails". 
The symbols used to identify the three tails are the
same as those used in Fig.~\ref{rrab} to plot the different light curves.}
\label{tail}
\end{figure}

The RRab class is richer in particularities than the RRc one. In the phase diagrams
there is an unique, narrow progression from 0.40~d to 0.55~d. After this
uniform behaviour, we see the $\phi_{i1}$ values diverge into
three ``tails". 
The Fourier \pqu parameters 
provide the clearest separation among tails (Fig.~\ref{tail}).
The separation is usually larger than 0.50~rad, while  the fits
of the light curves give a mean error on the \pqu parameter of 0.16, 0.14
and 0.08 rad for the upper, central and lower tails, respectively.
In a few cases, looking at a particular $\phi_{ji}-P$ plot,  it is not easy 
to decide which tail a star belongs to,
owing to the error on the phase parameter we are examinating. However, this task
can be done in a much more reliable way by 
considering all the $\phi_{j1}-P$ plots; in that way we take full advantage 
of the presence of 
the three tails in each plot. As proven by the three panels in Fig.~\ref{tail},
the same stars form the same type of tail, though with a different degree of clarity.
The upper and central tails are
merged together for $P<$0.55~d and cannot be separated;  the lower
tail originates at $P\sim$0.50~d. Moreover, the OGLE sample does not contain many stars
with $P>0.65$~d; therefore, this part is not well defined in each tail (see
Sects. 5.1 and 5.2).

\begin{figure}
\resizebox{\hsize}{!}{\includegraphics{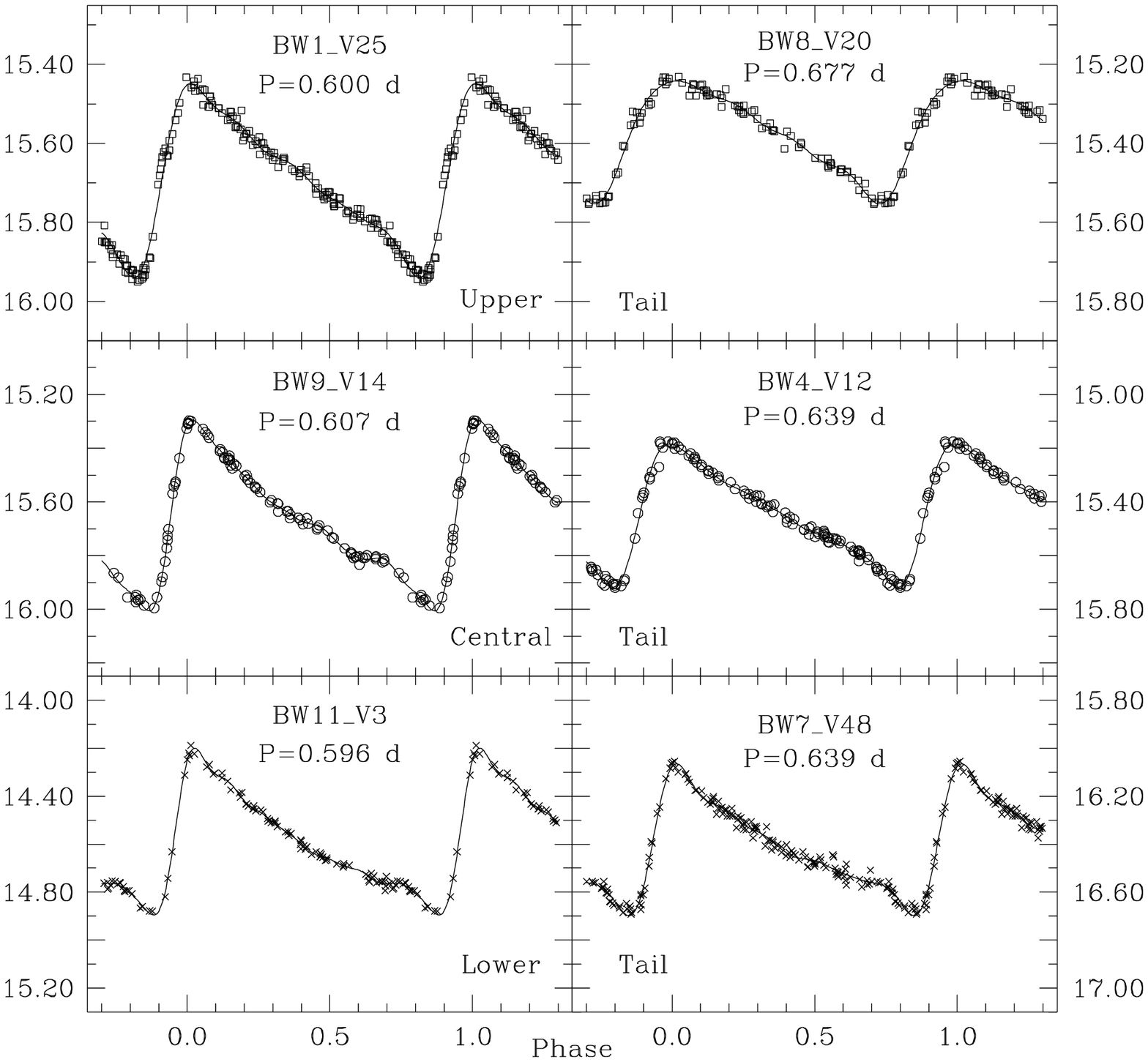}}
\caption[ ]{Light curves of different RRab stars belonging to the three
different tails visible in the phase Fourier plots. The upper panels (squares)
show a short-- and  a long--period RRab characterized by high values of the phase
differences (upper tail). The central panels (circles) show a short-- and
 a long--period
RRab characterized by medium values of the phase differences (central tail).
The lower panels (crosses) show a short-- and  a long--period RRab characterized by low
values of the phase differences (lower tail). The RR Lyr variables in each column have
approximately the same period and thus differences in the light curve shape
at the same period can also be assessed. Note also the differences in the amplitudes.}
\label{rrab}
\end{figure}

Once evidenced in the Fourier plots,  the differences 
can be evaluated also by a direct examination of the
light curves (Fig.~\ref{rrab}): 
\begin{enumerate}
\item The lower
tail is formed by stars having the steepest rising branches and  well
defined bumps before them; the maxima are sharp
and the descending branches are concave (Fig.~\ref{rrab}, lower panels). Moreover,
there are no great changes when the period increases (compare
the light curve of BW11\_V3, $P$=0.596~d, with that of BW7\_V48, $P$=0.639~d).
Stars belonging to this tail can be recovered down to $P$=0.523~d.
\item The light curves of the stars forming the central tail show a less steep ascending
branch; the descending branch has a sort of corrugated shape (Fig.~\ref{rrab},
middle panels; note that the points, not the fitting curve, are
describing it) for shorter periods (BW9\_V14,
$P$=0.607~d) and shows a tendency to become linear for longer
periods (BW4\_V12, $P$=0.639~d). Moreover,  the bump is wider
and less marked than in the lower tail. 
\item The upper tail is formed by 
light curves having the slowest rise to maximum (Fig.~\ref{rrab}, upper panels). 
The bump at minimum light is
visible at short periods (BW1\_V25, $P$=0.600~d), but disappears at longer ones
(BW8\_V20, $P$=0.677~d); moreover, the descending branch looks convex.
The amplitude is smaller than in the two previous cases and a 2$^{\rm nd}$--order
fit is often sufficient. 
Consequently, the $R_{21}-P$ plot clearly separates the 
stars belonging to this tail
from other RRab variables; in Fig.~\ref{r21} they
originate the points below 0.4 after $P$=0.60 d. 
\end{enumerate}

As the general
shape of the curves changes from one tail to another, all the phase parameters
are sensitive to differences, particularly the  higher--order ones,
since they should fit the bump at minimum and the
differences in skewness. This explains why the separation of the tails is
more conspicuous  in the \pqu$-P$ and $\phi_{51}-P$ plots than in the \pdu$-P$ 
and \ptu$-P$ plots.

\begin{figure}
\resizebox{\hsize}{!}{\includegraphics{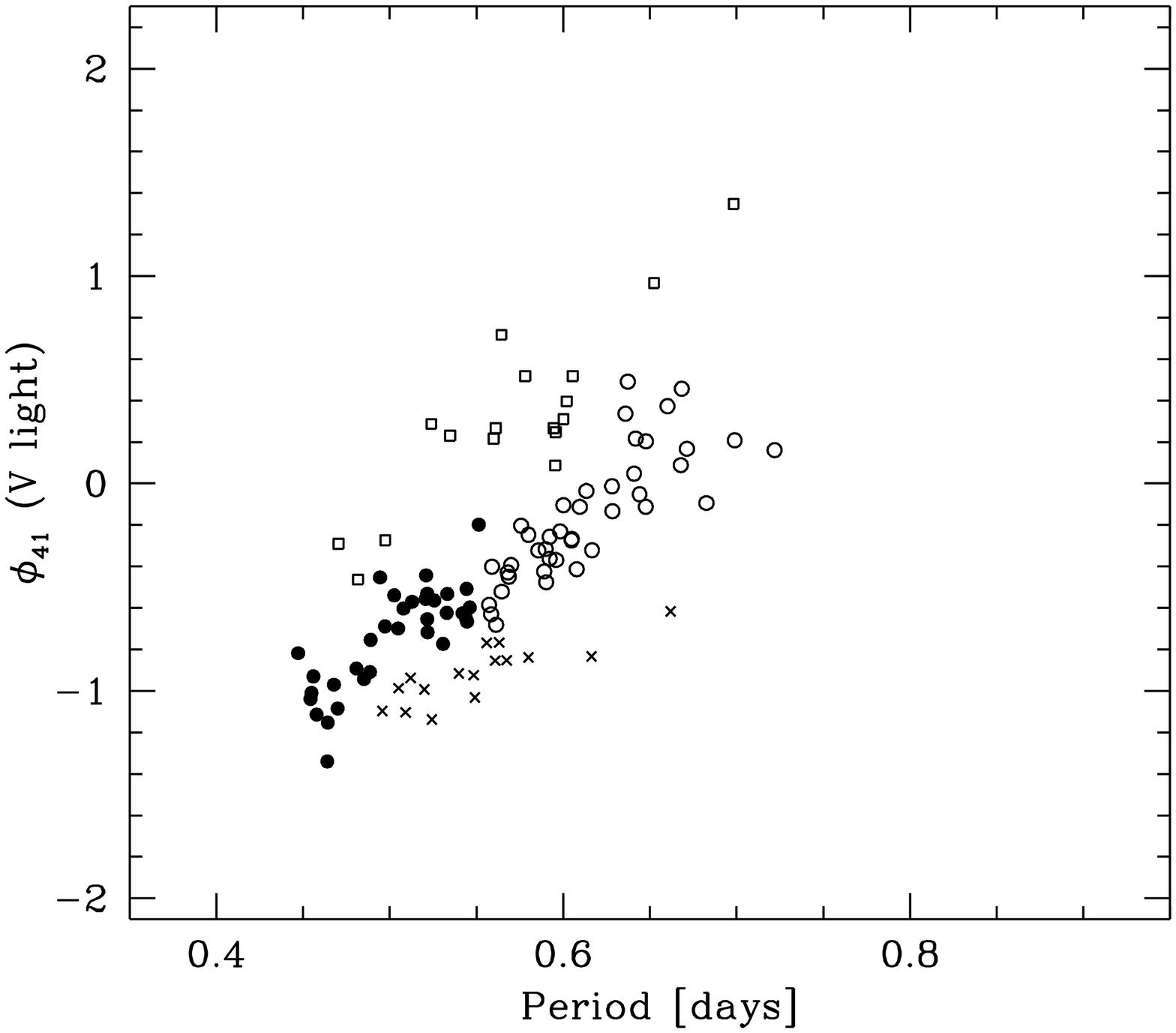}}
\caption[ ]{Fourier \pqu parameters of RRab stars in some globular clusters ($V$
light). 
The central tail (filled and open circles) is composed by stars from NGC~6362, M55, M9, M5, NGC~6934 and
M3. The upper tail (open squares) is composed by  all the stars from NGC~6171 and few stars 
from NGC~6362, M5 and
M3. The lower tail (crosses) indicates stars from M55 and NGC 6934.} 
\label{glob}
\end{figure}

\section{Discussion}

The complete re--analysis of the OGLE database 
gave us the opportunity to demonstrate the non--existence of the
unusual $\delta$ Scuti stars reported by Morgan et al. (1998c), to 
recognize the bimodal distribution of the \Rdu parameters in the HADS sample
as a contamination effect, to separate in a better way
the three classes of variable stars. Other aspects deserve further
discussion, also considering both the inputs needed by theoretical work and
the necessity of obtaining clear indications
from  monoperiodic stars in order to better understand multimode pulsators.

\subsection{Features in the progressions}
The Fourier parameters are confirmed to be a powerful discriminator between
the first overtone mode of RRc stars and the fundamental mode of RRab stars.
The same separation
is discernible, though with more difficulty, between the RRc stars and the
HADS stars, considered to be $F$ mode pulsators. These latter stars show
a changing slope at $P\sim$0.10~d; the possibility of  a presence of 1$O$
pulsators among short--period HADS is suggested by some sine--shaped light
curves and by a small group of stars having high values of \pdu (Poretti
1999). 

None of these progressions
shows a clear, abrupt change suggesting the presence of  a resonance.
We can speculate that the upper tail of the RRab changes
at $P=$0.64~d (Fig.\ref{tail}), as there are  
two stars (MM5-B\_V20 and BW10\_V95) showing a \Rdu value and a shape
typical of upper sequence stars, but a significantly lower \pqu value.
Moreover, the $\phi_{51}$ and $\phi_{61}$  parameters show a regular
progression, confirming that the two stars belong to the upper tail.
It will be interesting to verify the hint of a resonance involving
the 3rd overtone by means of a larger database.

\subsection{The three tails of the RRab progression}

Kov\'acs \& Jurcsik (1996, 1997) and Jurcsik (1998) have derived formulae that
estimate the absolute magnitudes, metallicities, intrinsic
colours and temperatures of RRab stars from the periods, amplitudes and phases.  
In particular, the [Fe/H] value is related to $P$ and \ptu.  Therefore, 
the differences in the progression naturally transpose into a difference
in metallicity. After transforming the phase differences values from $I$ to 
$V$ light using the equations provided by Morgan et al. (1998b), we indeed 
obtained a slightly different
mean metallicity: the 20 stars of the upper sequence yield [Fe/H]=--0.96$\pm$0.04,
the 17 stars of the central one --1.39$\pm$0.05 and the 9 stars of the lower one
--1.50$\pm$0.08. We note that they are strictly monoperiodic pulsators,
and the condition of regular light curves (Kov\'acs \& Kanbur 1998)
is then largely satisfied.

To investigate in more detail the possible
effect of the metallicity,
we revisited the \pqu$-P$ progression for the RR Lyr variables in NGC~6171
(Clement \& Shelton 1997; we omitted two unclear cases), NGC 6362
(Olech et al. 2001), M3 and M9 (Clement \&
Shelton 1999), M55 (Olech et al. 1999), M5 (Kaluzny et al. 2000a) and NGC 6934
(Kaluzny et al. 2000b). We transformed the phase values obtained from a Fourier
series in {\tt sin} into phase values  from a {\tt cos} series. The parameters
obtained in this way are plotted in Fig.\ref{glob} using the same symbols
as in Fig.~\ref{tail} to indicate the different tails. Note the sharpness
of the OGLE progressions compared with the progressions relative to cluster
stars; observations
carried out with different instruments 
and much more crowded fields can explain the larger scatter.

Most of these globular clusters show overlapping
central tails (NGC 6362, M9, M5 and NGC 6934),  
despite  a large spread in metallicities. However, looking at Table~\ref{gc}
we can see that the expected general distribution is respected: metal-rich
globular clusters have some or most stars on the upper tail
and none on the lower one, while the RR Lyr stars of the metal--poor clusters 
populate the lower tail. 
Note also that the two extreme cases (NGC 6171, metal--rich, 
and  M55, metal--poor) describe  the two progressions
well above and below the central one, respectively.
\begin{table*}
\caption[]{Distribution of the cluster RRab light curves
in the different segments (common part with $P<$ 0.50 d and three tails)
of the \pqu$-P$ diagrams.}
\begin{flushleft}
\begin{tabular}{lcrc rrrrrr}
\hline
\multicolumn{2}{c}{}&\multicolumn{1}{c}{}&&
\multicolumn{1}{c}{}&\multicolumn{1}{c}{}&
\multicolumn{3}{c}{Tails}&  \\
\cline{7-9}
\multicolumn{2}{c}{Glob. Clus.}&\multicolumn{1}{c}{[Fe/H]}&&
\multicolumn{1}{c}{N}&\multicolumn{1}{c}{RRab}&\multicolumn{1}{c}{Central}&
\multicolumn{1}{c}{Upper}&\multicolumn{1}{c}{Lower}\\
\multicolumn{2}{c}{}&\multicolumn{1}{c}{}&&
\multicolumn{1}{c}{}&\multicolumn{1}{c}{$P<$0.50 d}&\multicolumn{1}{c}{$P>$0.50 d}&
\multicolumn{1}{c}{}&\multicolumn{1}{c}{}\\
\noalign{\smallskip}
\hline
NGC 6362  & & --0.95     &&  18 & 11 & 3 & 4 & --  \\
NGC 6171  & & --1.04    &&   7 & -- & --& 7 & --  \\
\noalign{\smallskip}
M5        & & --1.29    &&  30 & 13 & 15& 2 & --  \\
M3        & & --1.47    &&  10 & 2  & 4 & 4 & --  \\
\noalign{\smallskip}
NGC 6934  & & --1.54    &&  31 &  7 & 10& --& 14  \\
M9        & & --1.72    &&   6 & -- & 6 &-- & --  \\
M55       & & --1.81    &&   6 &  1 & 3 &-- &  2  \\
\hline
\end{tabular}
\end{flushleft}
\label{gc}
\end{table*}

However, all the stars of the same cluster 
should have the same
metallicity and therefore the \pqu$-P$ plot of each cluster should show a
single tail, but in three cases we actually observe the co-existence of  
two kinds of tails (see also Tab.~1):
\begin{enumerate}
\item in M3 there is a
tendency for long--period RRab stars to bend toward higher \pqu values and
to merge in the upper tail (see Clement \& Shelton 1999);
\item in NGC 6934, the
RRab variables with 0.49 d$<P<$0.56 d seem to belong to the lower sequence
and the others to the central one;
\item in NGC 6362 there are some RRab stars with $P>$~0.60~d
belonging  to the  upper tail and other RRab belonging to the central tail.
\end{enumerate}
Therefore, it looks evident that only the general trend of a large
sample can be considered the fingerprint of the metallicity of the cluster,
not the Fourier parameters obtained from a few stars. 

In the OGLE database the sample is three times more numerous and the separation
between the tails looks better defined than in the Clement \& Shelton (1999, their Fig.~8)
sample. The RRab stars in $\omega$ Cen describe a sharp central tail
(Poretti 2001, in preparation); a 
comparison with  Fig.~\ref{tail} supports the hypothesis that 
the three OGLE stars at $P\sim$0.80~d belong  to this
tail.

The presence of the tails in the OGLE database indicates an additional
influence apart from metallicity on the light curve shape. 
We can conjecture that the shapes of the light curves are related 
to something  more general, such as the evolutionary phase, or a combination
of different physical conditions, since no obvious metallicity--based
segregation is expected for field RR Lyr stars. 

\subsection{RRc stars}
The slope in the \ptu$-P$ plot is steep. The relationships
given by Simon \& Clement (1993) allowed us to calculate some 
physical parameters. Morgan et al. (1998b) found a very homogeneous group, with
a few stars showing unusual metallicity.
We subdivided the RRc sample into two subsets,
the first (21 stars) with 0.25$<P<$0.30 d and the second (17 stars)
with 0.30$<P<$0.35 d. There is no difference in  luminosity (log$L$=
1.65$\pm$0.01 and 1.67$\pm$0.01, respectively) and in the relative helium abundance 
($Y$=0.29 everywhere) between the two subsets. On the other hand,
the RRc stars with the shorter period
seem  more massive (mean mass 0.57$\pm$0.02$M_{\sun}$ vs. 
0.50$\pm$0.02 $M_{\sun}$) and hotter (mean temperature $T$=7460$\pm$10~K
vs. $T$=7380$\pm$10 K), but the differences can be
considered marginal. However, this tendency is 
confirmed when  comparing the two RRc variables having $P<$0.25~d 
(0.56 and 0.57 $M_{\sun}$, 7726~K and 7649~K) with the two RRc stars 
having $P>$0.35~d (0.58 and 0.78 $M_{\sun}$, 7177~K and 7047~K). 

When considering different globular clusters we note that metallicity 
is driving the
differences in the physical parameters (the mean masses and mean luminosities
increase and the mean temperatures and helium abundances decrease with decreasing
cluster metal abundance, Clement \& Shelton 1997). In the OGLE database and
in the same globular cluster (i.e. under the same metallicity), the  
RRc variables  with shorter periods are 
the more massive and hotter stars (see  V2 and V11
in M55, Olech et al. 1999, or V14 and V28 in NGC6362, Olech et. al. 2000)

This tendency needs to be better fixed   
by a more robust observational sample, which consider also RRc stars with
$P<0.25$~d (i.e. the extension of the RRc class we introduced here) and
RRc stars with $P>0.35$~d.

\subsection{The comparison with the other field RR Lyr stars} 
Simon \& Teays (1982) made the first Fourier decomposition of field RR Lyr
stars, including both RRab-- and RRc--types (70 stars). Subsequent 
papers did not further discuss their conclusions. As regards RRab stars,
Simon \& Teays  claimed  evidence for a trend, a sort of Cepheid--like progression,
for $P>$~0.55~d. This progression is masked by a large scatter, not
observed in the OGLE sample. As a consequence, these authors 
missed the opportunity to note that the progression originated
earlier, around $P\sim$0.45~d and to note the different tails.
The reason for such a  scatter 
is unclear, since observational errors or uncertainties on
the parameters are not able to explain it fully. 

Simon \& Teays also emphasized the unusual case constituted 
by XZ Cet ($P$=0.8231~d),
by far their longest--period star (gap from 0.75 to 0.82~d). The OGLE sample 
fills such a   gap only partially (four stars clustering around 0.77~d) and 
individuates a longer--period variable, MM7-B\_V12 ($P$=0.8812~d). 
Considering the whole sample of long--period RRab stars,
the general pattern seems to be a continuous progression, not interrupted by
any discontinuity, as suggested by Simon \& Teays (1982). We also note  
that these long--period stars are not found in the globular clusters we
considered above:
in Fig.\ref{glob} the longest period is 0.72~d. As a matter of fact,
they seem to characterize
metal--rich globular clusters, as NGC 6388 and NGC 6441 (Pritzl et al. 2000).

\section{Conclusions}
The Fourier coefficients with their error bars are reported in
a machine readable form in Tab.~3 (HADS), Tab.~4 (RRc) and Tab.~5 (RRab).
Moreover, an atlas of the light curves, least--squares fit and
Fourier parameters  is available as a ps-file from the author.

The step--by--step analysis we made allowed us to handle a sample
characterized in an unambiguous way: no double--mode or Blazhko
pulsator, no under-- or over--evaluation of the  significance of
the Fourier components. Consequently, clear outputs have been obtained.

The utility of the Fourier decomposition as a classification tool is
confirmed by the segregation of the phase and amplitude parameters in
well--defined loci for each class of variables.
The use of this technique as a 
link between observable quantities  and stellar physical parameters is
demonstrated by the enhanced separations within the same class, as examplified
by the
three tails in the RRab phase parameter plots. This way of performing the
analysis when managing limited samples should give some warnings
about the automated analysis of very large samples. A good trade--off
could be the introduction of  sophisticated checks at intermediate 
steps.

In a more general way, the exploitation of the large number of
light curves supplied by the microlensing projects should be carried out 
in a careful way, since it can really provide us with the possibility to 
sound the stellar interiors.

\begin{acknowledgements}
Useful comments on a first draft of the manuscript were made
by Pawel Moskalik, Siobahn Morgan, Elio Antonello, Geza Kovacs,
Christine M. Clement, Giuseppe Bono 
and the referee, Matthew Templeton.
A part of this work was done during a stay at the Copernicus
Astronomical Center in Warsaw and the author wishes to thank
P.~Moskalik and W.~Dziembowski for their warm hospitality.
An early short visit to Siobahn Morgan at University of Northern Iowa was
very helpful in planning this work.
The author wishes also to thank
J.~Vialle for checking English form.
\end{acknowledgements}

\end{document}